\documentstyle[11pt,newpasp,twoside,epsf]{article}
\def\edcomment#1{\iffalse\marginpar{\raggedright\sl#1\/}\else\relax\fi}
\reversemarginpar

\begin{document}
\title{Grids of synthetic spectra for the GAIA mission}
 \author{Toma\v{z} Zwitter}
\affil{University of Ljubljana, Dept.\ of Physics, Jadranska 19,
1000 Ljubljana, Slovenia, tomaz.zwitter@fmf.uni-lj.si}
\author{Ulisse Munari}
\affil{Osservatorio Astronomico di Padova -- INAF, Sede di Asiago, I-36012
Asiago(VI), Italy, munari@pd.astro.it}
\author{Fiorella Castelli}
\affil{Osservatorio Astronomico di Trieste, via Tiepolo 11, 34131 Trieste, 
Italy, castelli@ts.astro.it}

\begin{abstract}
Two sets of grids of Kurucz ATLAS~9 model spectra are presented. 
The first one covers the GAIA spectral interval, while the second 
one which is still being computed covers the whole range from 
2500 -- 10500~\AA. Both grids include parameters needed for realistic 
spectral simulation, e.g.\ stellar rotation and are computed at 
$R=20000$ and lower spectral resolutions. They can be 
useful for preparation of the GAIA and RAVE missions and also for a  
general radial velocity correlation work.
\end{abstract}

\section{Introduction}

GAIA mission will observe an unprecedented number of stars and will 
set a new standard reference in stellar photometry and spectroscopy. 
Preparations for such a mission need to build on a vast body of 
observed and synthetic spectra. On one hand the spectra should 
cover the baselined GAIA spectral interval (8490 -- 8750 \AA) at 
different resolutions and spectral samplings. But the wavelength 
range of ground-based Echelle spectrographs is much wider. Synthetic 
spectra covering the whole range from the near-UV to the near-IR 
can be useful for general radial velocity correlation work. They 
can also help to optimize the scientific output of broad and 
narrow-band photometric filters aboard GAIA (Jordi 2002).


Observed spectra of normal stars in the GAIA spectral interval 
were presented by Munari \&\ Tomasella (1999). The first grid of 
Kurucz synthetic spectra was published by Munari \&\ Castelli (2000) 
and Castelli \&\ Munari (2001). Here we report on the ongoing effort 
to expand this theoretical grid using the same sets of abundancies and 
atomic constants. 

\section{Spectra in the GAIA spectrograph wavelength range}

Synthetic spectra are needed as templates for measurement of 
radial velocity by correlation techniques (Zwitter 2002). But 
another use may be less obvious and equally important. Virtually all 
spectra obtained by GAIA will suffer from overlaps of faint spectral 
tracings of neighboring stars (Zwitter \&\ Henden 2002). Most of the 
overlappers will be too faint to recover their spectrum from GAIA 
observations. So one will have to rely on photometric classification 
of the overlapping stars and a proper synthetic spectra database to generate
a combined spectrum of the overlapping background stars, subtract it 
from the studied spectrum and finally analyze it (Zwitter 2002a). One 
could therefore refer to the synthetic spectra database as a critical 
part of the data reduction code, as no information could be extracted 
from the spectrograph without subtraction of the calculated background 
signal.

Most of the GAIA stars will be normal stars of spectral types G and K
(Zwitter \&\ Henden 2002). Thus Kurucz ATLAS~9 stellar atmosphere
models can be used as an initial approach to synthetic grid calculation. 
They are also adequate for instrument and reduction procedure planning. 
We note that non-LTE effects, presence of dust at low temperatures and 
other peculiarities need to be taken into account in certain 
cases (Hauschildt 2002).

The primary goal of the GAIA spectrograph is to measure radial velocities.
Moreover, the spectra can be used to calculate or confirm the effective 
temperature, gravity and metallicity of observed stars and so supplement 
the photometric results. Finally, in the case of bright targets and 
provided that spectral resolution is high-enough, the GAIA spectra can 
also be used to obtain abundances of individual elements, measure 
stellar rotation and explore spectral peculiarities 
(Munari 2002; Thevenin 2002; Gomboc 2002). 

\begin{table}
\caption{
Grid of spectra in the GAIA spectrograph wavelength range. Some models 
were calculated also for an $\alpha$-enhanced composition.
}
\begin{tabular}{lllll}
&&&&\\
\tableline
&&&&\\ [-5pt]
parameter&min&max&step&comment\\
\tableline
&&&&\\ [-5pt]
$T_{eff}$        & 3500~K & 50000~K & 250~K & larger steps for $T_{eff}>10000$~K\\
$\log g$         & 0.0    & 5.0     & 0.5   & till 4.0 for hot stars  \\
$[$Z$/$Z$_o]$    &-3.5    & +0.5    & 0.5   &                             \\
$v_{rot} \sin i$ & 0 km$/$s& 500 km$/$s&    &14 values for O-F stars\\
                 &              &                & &11 values $\le 100$ 
		                                   km$/$s for G-M\\
$R$              & 5000         & 20000     & & 5000, 10000, 20000\\
 [4pt]
\tableline
\end{tabular}
\end{table}

With these goals in mind we generated a grid of synthetic spectra based 
on the Kurucz ATLAS~9 models computed for a microturbulent velocity of 
2 km~s$^{-1}$, a mixing-length convection with parameter $l/H=1.25 $ and no 
overshooting. The spectra were computed with the SYNTHE code from Kurucz and 
cover the spectral range 8490-8750 \AA\ of the GAIA RVS spectrograph. Details
on the database computation will be published elsewhere (Zwitter et al.\ 
2003, in preparation). Table 1 gives the ranges of the spectral grid 
parameters. Apart from the basic quantities ($T_{eff}$, $\log g$ and 
[Z$/$Z$_o$]) we include additional dimensions, i.e.\ rotational velocity 
($v \sin i$) and spectral resolution ($R$). Thus the spectral tracings 
are ready to be included in realistic GAIA simulations. Altogether the 
database now consists of $\sim 2 \times 10^5$ spectra. 

\begin{figure}
\plotfiddle{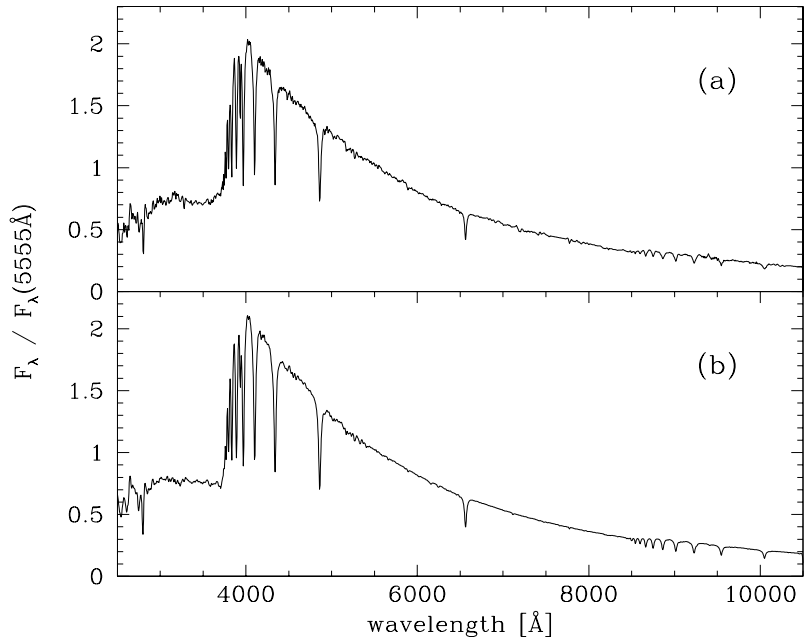}{7.9cm}{270}{45}{44}{-170}{250}
\caption{(a) Observed spectrum of an A5~V star  
at a resolution $R=500$ sampled with 5\AA\ steps (adapted from Pickles 1998). 
(b) Synthetic spectrum of a non-rotating star with
$T_{eff} = 8250$~K, $\log g = 4.0$ and solar composition resampled to 
the same resolution and sampling.  
}
\end{figure}

\section{Spectra from the near ultraviolet to the near infrared}

Recent advances in storage space and computing power permit a calculation
of a database of spectra covering the 2500~\AA\ to 10500~\AA\ wavelength range.
It was computed with Kurucz's codes at a resolution of $R=500000$ and degraded 
to $R=20000$, i.e.\ the one typical for Echelle spectrographs. 
The spectra can be used for general radial velocity correlation work. 
Moreover the radial velocity solutions based on the GAIA spectral interval 
can be compared to the velocities extracted from the whole wavelength range. 
Such spectra can be also degraded to lower resolutions and so of 
interest for optimization of the scientific 
output from the GAIA's narrow and broad band photometric observations.

In Table 2 we report on the grid calculated so 
far, which covers the most common stars to be observed by GAIA. 

\begin{table}
\caption{
Grid of spectra with $R=20000$ for the 2500~\AA~$< \lambda <$ 10500~\AA\ range.
They can be degraded to any lower resolution.}
\begin{tabular}{lllll}
&&&&\\
\tableline
&&&&\\ [-5pt]
parameter&min&max&step&comment\\
\tableline
&&&&\\ [-5pt]
$T_{eff}$        & 5250~K & 10000~K & 250~K & \\
$\log g$         & 0.0    & 5.0     & 0.5   &   \\
$[$Z$/$Z$_o]$    &-3.5    & +0.5    & 0.5   &                             \\
$v_{rot} \sin i$ & 0 km$/$s& 500 km$/$s&    &14 values for O-F stars\\
                 &              &                & &11 values $\le 100$ 
		                                   km$/$s for G-M\\
 [4pt]
\tableline
\end{tabular}
\end{table}

\begin{figure}
\plotfiddle{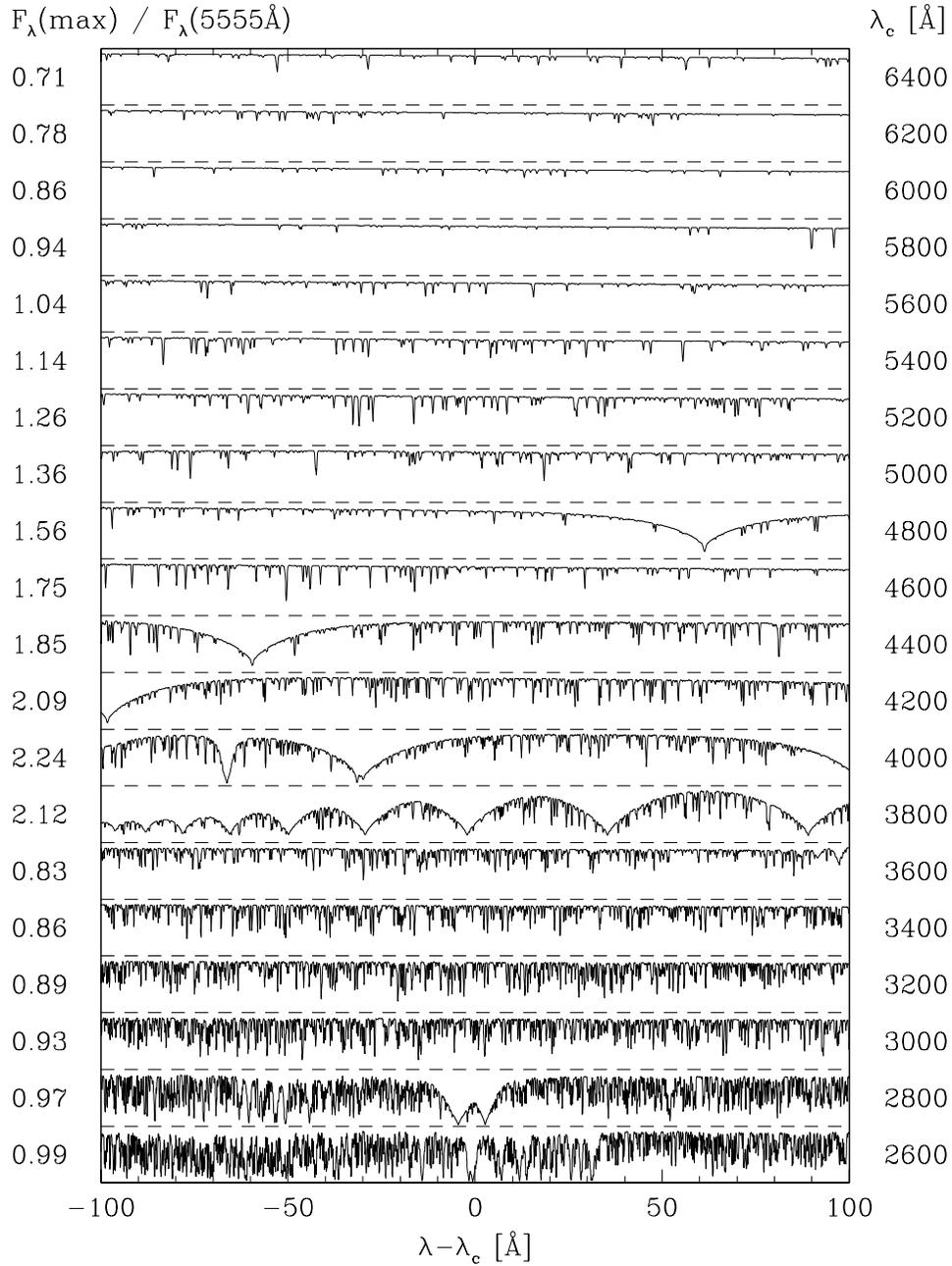}{17.0cm}{0}{67}{70}{-205}{-30}
\caption{Blue part of the synthetic spectrum with $T_{eff} = 8250$~K, 
$\log g = 4.0$, solar composition and resolution $R=20000$. Numbers 
on the right give the central wavelength and those on the left the maximum
flux in each window. The lower limit of the ordinate is always zero.
}
\end{figure}

\begin{figure}
\plotfiddle{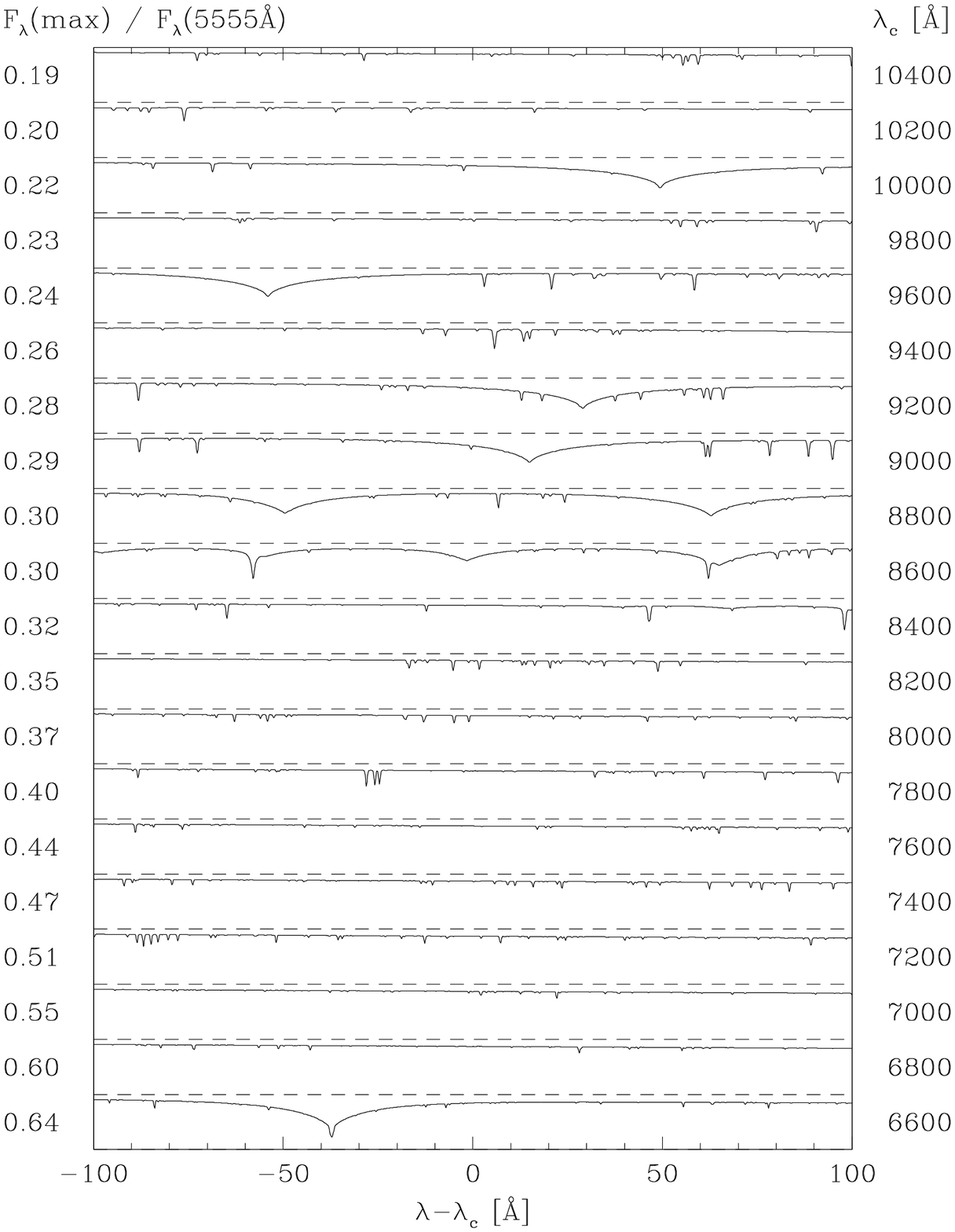}{15.7cm}{0}{67}{70}{-205}{-30}
\caption{Red part of the synthetic spectrum from Fig.~2.}
\end{figure}

Figure 1 compares an observed and synthetic spectrum of an A5~V star. 
The observed spectrum was obtained by combining several observed 
standard star spectra (Pickles 1998) and has a resolution of 500 sampled 
at 5~\AA\ wavelength bins. The calculated spectrum was resampled to 
the same resolution and sampling. It was not optimized to match the 
observed one, still the differences are small. Figures 2 and 3 show 
the same calculated spectrum at full resolution ($R=20000$).

\section{Conclusions}

The grid of synthetic spectra based on the Kurucz models and covering the 
GAIA spectral interval 8490-8750 \AA\ is almost complete. The computation 
of the other grid with full wavelength coverage is still being done. We note 
that these spectral databases should be useful for the GAIA as well as 
the forthcoming RAVE missions, and also for general tasks of 
radial velocity correlations.

\end{document}